# Visualizing the Process of Process Modeling with PPMCharts[1]

(regular paper)

Jan Claes[1], Irene Vanderfeesten[2], Jakob Pinggera[3], Hajo A. Reijers[2], Barbara Weber[3] and Geert Poels[1]

[1] Ghent University, Belgium
{jan.claes, geert.poels}@ugent.be
[2] Eindhoven University of Technology, The Netherlands
{i.t.p.vanderfeesten, h.a.reijers}@tue.nl
[3] University of Innsbruck, Austria
{jakob.pinggera, barbara.weber}@uibk.ac.at

**Abstract.** In the quest for knowledge about how to make good process models, recent research focus is shifting from studying the quality of process models to studying the process of process modeling (often abbreviated as *PPM*) itself. This paper reports on our efforts to visualize this specific process in such a way that relevant characteristics of the modeling process can be observed graphically. By recording each modeling operation in a modeling process, one can build an event log that can be used as input for the *PPMChart analysis* plug-in we implemented in ProM. The graphical representation this plug-in generates allows for the discovery of different patterns of the process of process modeling. It also provides different views on the process of process modeling (by configuring and filtering the charts).

## 1      Introduction

"A picture is worth a thousand words." This phrase is believed to originate from an Asian verb and advocates the use of visualization. The actual value of the graphical representation of models, however, is heavily influenced by its understandability [1, 2]. In our research we look for determinants of the modeling process that influence the understandability of the process modeling result (i.e., a process model) [3]. The visualization presented in this paper is developed to support our research to the process of process modeling. The *process of process modeling* is the course of action taken by the modeler to create/design/construct a (business) process model consisting of start and end event(s), activities, gateways, edges, etc. Such a process model artifact is created by a stepwise design process, e.g., first putting a start event on the canvas, then an activity, then an arc connecting the start event and the activity, etc.

---

[1] The final publication is available at Springer via http://dx.doi.org/10.1007/978-3-642-36285-9_75

To get insights into how process models are constructed, we searched for a technique to visualize the process of process modeling based on a log of recorded modeling actions, such that one can *quickly* obtain *in-depth* insights in the visualized process. In earlier research we used Modeling Phase Diagrams [4] to focus on high-level, less detailed process characteristics: i.e., modeling phases. A modeling phase (e.g., reconciliation phase) summarizes a set of 'equivalent' events in the process (e.g., move_node, create_edge_bendpoint) as one phase of which a modeling phase diagram shows only the duration and number of constructed model elements. For the more extensive representation described in this paper, we drew inspiration from the *Dotted Chart Analysis* plug-in of the process mining framework ProM [5]. This technique provides much detail and at the same time lets the user take a helicopter view on the visualized process. We extended the existing plug-in to solve three concrete issues when representing an instance of the process of process modeling:

(i) the process of process modeling has a fixed set of possible events and therefore we mapped these events on *fixed colors* used as in the visualization, which allows for visually comparing different charts,
(ii) to establish a clear link between the visualization of the process and the constructed process model, we provided *two extra sort options* that are based on the execution order of the modeling elements in the process model,
(iii) we added the possibility to *filter* certain modeling operations in order to be able to take different, more abstract views on the same modeling process.

The visualization technique presented in this paper allows us to zoom in on the separate operations of the construction of the model. At the same time, it provides an overview of the entire modeling process. In this way it enables to obtain graphically a fast but detailed impression of how a process modeling effort was conducted. The acquired insights facilitate the study of the process of process modeling.

The structure of the paper is as follows. Section 2 presents related research. Section 3 describes the necessary data for our visualization, the visualization itself, and the implemented tool support. Section 4 illustrates the use of the tool by describing how it was used to discover patterns in process model construction. Finally, Section 4.3.5 concludes with a discussion and an overview of future research plans.

## 2    Related Research

There is a wide body of literature that focuses on the *quality* of process models [6–10]. Mostly, the process model is considered in these papers as a given, complete, and finished artifact. Other literature reports on research on *methods* for business process modeling (e.g., [11] provides a comparative analysis between different techniques for business process modeling and contains an extensive list of related papers on process model notations). Recently, approaches are emerging that aim to connect the previous two topics: In what way the used (in)formal modeling method relates to the properties (e.g., quality) of the outcome: a process model? In this context, various authors refer to the construction of a process model as *the process of process modeling* [12–15], a term often abbreviated as *PPM*.

Crapo et al., report on research about the process of modeling in terms of visualization (i.e., they focus on graphical modeling) [16]. However, we are not aware of other research about the visualization *of* the process of process modeling. In [4] the possibilities of representing different phases of process modeling in Modeling Phase Diagrams were examined. Three specific phases are distinguished: comprehension, modeling and reconciliation. The visualization described in [4] differs from the one presented in this paper by the level of abstraction: While a Model Phase Diagram abstracts from individual modeling operations, our representation shows all recorded operations and all present information about the operations (e.g., timing aspects, model element type, and order in the model).

The ultimate goal of our research is to improve knowledge about how to make correct and more understandable process models. Other research that provide guidelines or techniques for improved business process modeling includes Seven process modeling guidelines (7PMG) [17] and Guidelines of Modeling (GoM) [18].

## 3  Visualizing the Process of Process Modeling with PPMCharts

We called our visualization of the process of process modeling a *PPMChart*. Section 3.1 explains which data is used for construction of the charts, Section 3.2 describes the properties of PPMCharts, and Section 0 centers on the developed tool support for generating PPMCharts.

### 3.1  Data collection

In order to visualize the process of process modeling, we use a record of modeling operations done by the modeler, and recorded by our experimentation tool, during the modeling process. The list of possible operations we consider is given in Table 1. For instance, the operation of creating an activity in the process model is logged by an event "CREATE_ACTIVITY". In our analysis and experiments we build on a subset of the BPMN notation that can be used for the modeling. This subset was selected to correspond with the supported notation of our experimentation tool (see Section 4.1) and consists of six of the ten most used elements of BPMN according to [19]: *start* and *end event*, *activities*, *XOR* and *AND gateways*, and *edges*.

Besides *creation* of these model elements, the visualization also includes changes in the model. Activities, events and gateways can be *moved* over the canvas or *deleted*. Edges can be *deleted* or *reconnected* (which we categorized as a deletion and creation), an edge can be rerouted through *creation*, *movement* and *deletion* of bendpoints, and the label of an edge can be *moved*. Finally, activities and edges can be *named* or *renamed*. Note, that for the rest of the paper we assume only these modeling operations as part of the modeling process (according to the recorded operations of the experimentation tool), but our approach can easily be adapted for other modeling operation sets.

Table 1. Operations in the construction of a process model

| Create | Move | Delete |
|---|---|---|
| CREATE_START_EVENT | MOVE_START_EVENT | DELETE_START_EVENT |
| CREATE_END_EVENT | MOVE_END_EVENT | DELETE_END_EVENT |
| CREATE_ACTIVITY | MOVE_ACTIVITY | DELETE_ACTIVITY |
| CREATE_XOR | MOVE_XOR | DELETE_XOR |
| CREATE_AND | MOVE_AND | DELETE_AND |
| CREATE_EDGE | MOVE_EDGE_LABEL | DELETE_EDGE |
| RECONNECT_EDGE (**) | CREATE_EDGE_BENDPOINT (*) | RECONNECT_EDGE (**) |
|  | MOVE_EDGE_BENDPOINT (*) |  |
|  | DELETE_EDGE_BENDBPOINT (*) |  |
| **Other** : NAME_ACTIVITY, RENAME_ACITIVTY, NAME_EDGE, RENAME_EDGE | | |

(*) create, move and delete edge bendpoint were considered as moving an edge

(**) reconnect edge was considered as deleting and creating an edge

### 3.2    Visualization with PPMChart

The collected data about consecutive operations in the process of process modeling for one modeler are used to construct a PPMChart (see Fig. 1). The horizontal axis represents a time interval of one hour. Vertically, each line represents one element of the model as it was present during modeling. Each dot represents one operation performed on the element; the color of the dot represents the type of operation: create (green), move (blue), delete (red) and (re)name (pink). The elements are vertically sorted by the time of their first operation; the first operation performed on each model element is its creation. The dots are aligned to the right such that the last operation performed by the modeler is shown to occur at the end of the one hour interval. shows the process model resulting from the process model construction process that is visualized in Fig. 1.

### 3.3    Tool support

We implemented a plug-in for the popular process mining framework ProM[2]. The input for most plug-ins in this tool is an event log. The file format for event logs for ProM is xml based and follows a certain hierarchical structure: A *process* consists of *traces* and each trace is a collection of *events*. The process, traces and events can have attributes (e.g., a time stamp). In our case *modeling operations* correspond with events and the operations on the same model element are bundled in one trace. We expect the *names* of events to correspond with the operations in Table 1. Further, each event has an attribute *id* that corresponds with the name of its trace. It can be seen as the unique identifier of the modeling element.

---

[2] For information and download we refer to http://www.promtools.org.

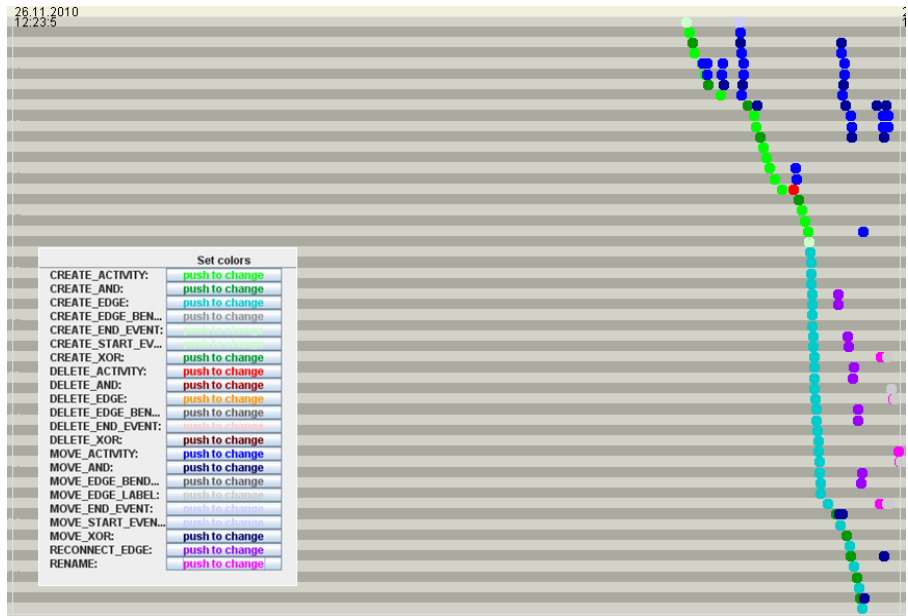

**Fig. 1.** Visualization of the operations in the creation of one model by one modeler.[3] The operations creating nodes (activities and gateways) are depicted as green dots; the creation of edges is depicted in light green/blue, the moving of model elements in darker blue, the deletion in reddish colors and the reconnect or renaming in pink/purple. The first line for instance shows the creation of the start event, followed by a move of the start event on a later time.

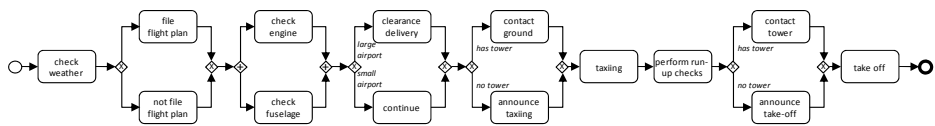

**Fig. 2.** Process model as result of the modeling process in Fig. 1.[2]

Our *PPMChart Analysis* plug-in (see Fig. 3) is an adapted version of the existing *Dotted Chart Analysis* plug-in [5]. In the middle, the PPMChart is presented. At the left-hand side, the view can be configured. At the right-hand side the user can customize the view by filtering on specific operations or elements: The top part represents a small view on the unfiltered PPMChart. Below, one can choose to hide specific element types (e.g., hide edges), hide specific operations (e.g., hide (re)name operations), or hide elements with a specific operation (e.g., hide deleted elements).

---

[3] High resolution graphs are available from http://www.janclaes.info/papers/PPMViz.

6      **Jan Claes, Irene Vanderfeesten, Jakob Pinggera, Hajo A. Reijers,** Barbara Weber and Geert Poels

For users that are familiar with the features of the Dotted Chart Analysis plug-in, we have chosen to keep the options of that plug-in in our implementation at the left-hand side of the window (see [5] for more information).

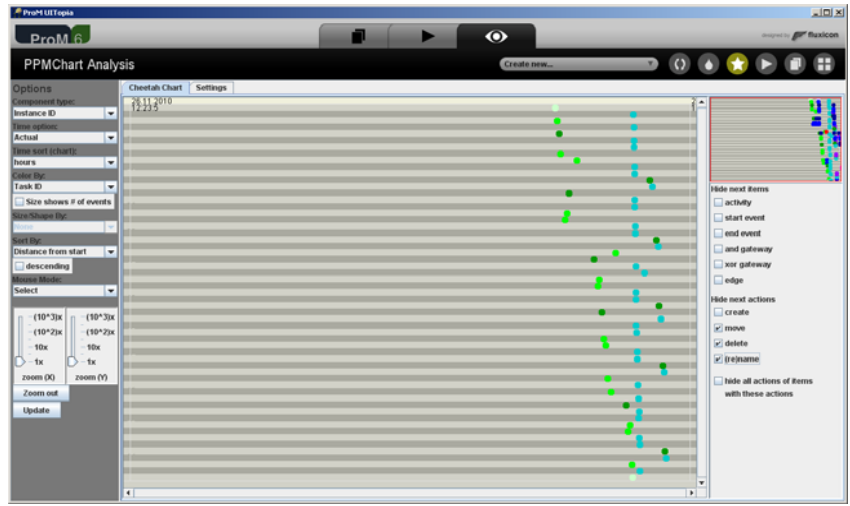

**Fig. 3.** Screenshot of the PPMChart window in ProM.[2]

To facilitate relating the different lines of the chart to the proper model elements in the process model, we also provided an additional *sort* option. This option sorts the lines of the chart from top to bottom according to the execution order of the model elements in the process model from start to end. Fig. 3 shows an example for the same case as for Fig. 1 and  but sorted in a different way. The green dots from top to bottom in the chart correspond to the model elements in the execution order in the process model (in Fig. 2 graphically from left to right). There are two variants of the additional sort option:

- The *Distance from start* sort option focuses on *execution* order. The value for nodes, events and gateways is defined as the sum of the lengths of the edges in the model, for the longest non-iterative path from that model element to the start event. The value for an edge is the average of the values for the source and target node of that edge.
- The *Create order from start* sort option centers on the *creation* of model elements from start to end and therefore it sorts an edge not between the nodes it connects, but after these nodes (an edge can only be created after both nodes exist). We use the same value for nodes, events and gateways, but for edges the value is one more than the maximum of values of source and target node.

Note that these calculations are highly influenced by the layout of the process model. This means that we adopt the modeler's view on the process in the visualizations.

For the example shown in Fig. 3, we selected 'Sort by Distance from start' and show only create operations (*start/end event* creation operations are colored light green, *activity* creations are bright green, *gateways* are dark green, *edges* are blue-green). It can be observed that the modeler first creates events, nodes and gateways in an order from start to end. Next, edges were created to connect the model elements. Finally, some extra nodes and edges were added. In the top right panel, all operations are displayed: Many intermediate move operations (blue dots) can be noticed.

## 4 Application of the PPMChart Visualization

Section 4.1 describes an experiment to which we applied our visualization technique to discover underlying patterns in process model construction. Section 4.2 describes how we used the experimental data to evaluate the correctness of our visualization technique. Section 4.3 lists a number of discovered patterns to demonstrate the usefulness of our technique for analyzing process modeling execution data.

### 4.1 Experiment design

In order to test the correctness and usefulness of our visualization and the implementation, we used data from our experiments on the process of process modeling, supported by Cheetah Experimental Platform (CEP)[5]. This platform instruments a basic process modeling editor to record each user's modeling operations along with the corresponding time stamps in an event log. Table 1 summarizes all recorded operations.

The experiment was conducted in November 2010 with 103 students following a graduate course on Business Process Management at Eindhoven University of Technology. Their task was to create a process model in BPMN from an informal description[6] while their process of process modeling was recorded in CEP. In the development of CEP, the developers decided to use a subset of BPMN (that we adopted for our research) without providing sophisticated tool features. The reasoning behind this was to not get the modelers confused or overwhelmed with tool aspects [16].

### 4.2 Evaluation of correctness: Replay the modeling process

By capturing all of the described interactions with the modeling tool, we are able to replay a recorded modeling process at any point in time, without interfering with the modeler. This allows for observing how the process model unfolds on the modeling canvas [4]. The evaluation of correctness was done by means of comparing the properties of the dots in the visualization with the observed properties of the steps in the replay. The test data concerned the process modeling efforts of the 103 students.

---

[5] For detailed information and download we refer to http://www.cheetahplatform.org.
[6] The case description is available at http://www.janclaes.info/papers/PPMViz.

8      **Jan Claes, Irene Vanderfeesten, Jakob Pinggera, Hajo A. Reijers,** Barbara Weber and Geert Poels

### 4.3   Demonstration of usefulness: Discovered properties and variations of the Process of Process Modeling

This section demonstrates the usefulness of the visualization by providing five concrete examples of properties of the process of process modeling that were derived from an analysis of the PPMCharts.

#### 4.3.1   General dimensions: Number of created model elements and duration of the modeling process

First, as one would expect from a visualization of the process of process modeling, the size and modeling time dimensions of the process can easily be compared between charts. Fig. 4a shows a variant where more modeling elements (91 elements) were created than the one in Fig. 4b (32 elements). Because operations for deleting or moving elements are included too, this provides more information than what can be obtained by comparing the resulting process models. Comparison of Fig. 4b and Fig. 4c reveals the difference in modeling time between two sessions (distances in the charts can be compared because the width of a PPMChart is always one hour).

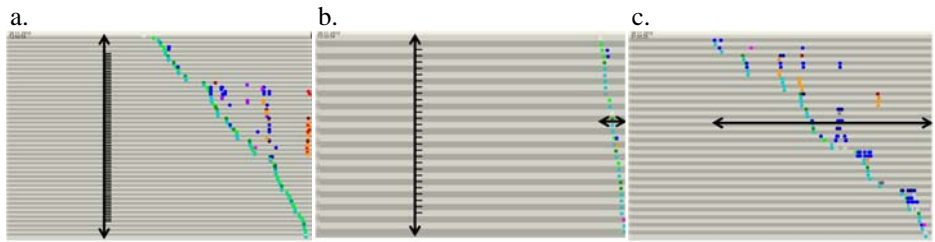

**Fig. 4.** Number of created model elements and duration of the modeling process.[2]

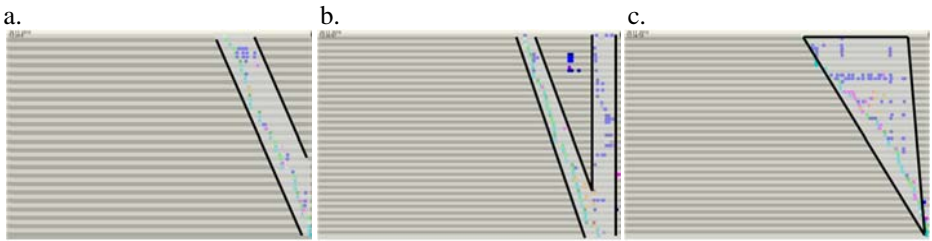

**Fig. 5.** Order of creating and changing model elements (sorted by First event).[2]

#### 4.3.2   Order of modeling operations (create, move, delete, (re)name)

Besides the general dimensions of the modeling process, one can also zoom in on the relative positions of the dots in the chart.  shows, for example, the modeling process of three modelers that moved some model elements after creation.  shows an example where elements were moved not long after their creation, the modeler of b performed most moves after *all* elements were created, while c shows a session in which move operations occurred during the whole modeling process.

Instead of using complex data mining techniques (e.g., Activity Clustering) to reveal these differences, the PPMChart allows for a quick visual detection of graphical patterns that provide the same information. The discovered modeling patterns can, for example, be used to identify different modeling styles.

### 4.3.3 Order of creation of elements (events, nodes, gateways, edges)

The filtering option enables to take different views on the process. By filtering the operations, one can, for example, focus on the order of creation of elements. Fig. 6 shows two extreme examples. The modeler of Fig. 6a first created the events, nodes and gateways and only afterwards the model elements were connected by edges. The modeler of Fig. 6b first created only events and nodes and afterwards the gateways and edges were created.

Even if the process would consist of hundreds of modeling operations, this could be quickly observed in our visualization. Arguably, it would take much longer to discover this from the pure data or by replaying individual models. The discovered patterns can, for example, be used to identify different modeler profiles.

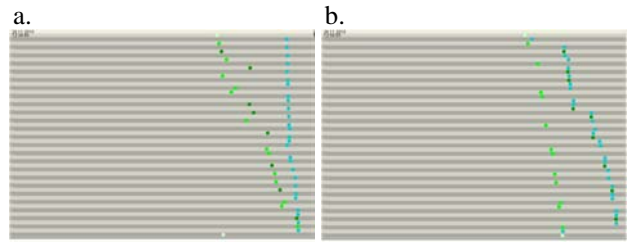

**Fig. 6.** Order of creation of elements (filtered view: only create operations).[2]

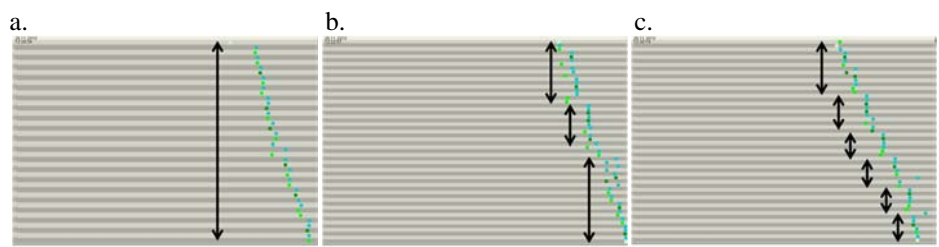

**Fig. 7.** Chunked process modeling (sorted by Distance from start, only create operations).[2]

### 4.3.4 Chunked process modeling

Because of the limited capacity of the human working memory [20], most modelers work in chunks, i.e., they work on calculable pieces of the model, one at a time. In the charts this can be observed in the form of clear pauses between the operations. In a the modeler seems to construct the whole model in a continuous way: We only observe a quite large pause after the creation of the start event. In b one could distinct three clear chunks of modeling operations. The modeler of c pauses after completion of each block (i.e., all elements between a set of matching split and join gateways, see Section 0).

The length of a period without operations that should be defined as a pause might differ between modelers because of the difference between slower and quicker modelers. However, the decision of which gaps are clear pauses seems to be easy when based on the graphical representation at hand.

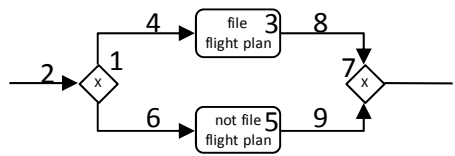

**Fig. 8.** Example of a 'block' in a model.[2]

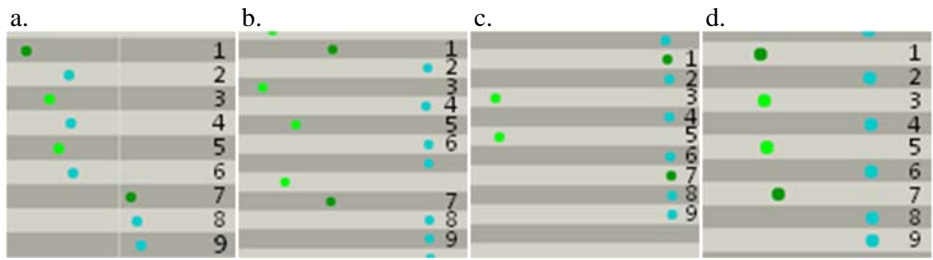

**Fig. 9.** Order of constructing blocks. The numbers correspond with the numbers in Fig. 8. (sorted by Create order from start, only create operations).[2]

### 4.3.5   Working in blocks: order of constructing blocks

The notion of a 'block' appears to be very relevant in the study of the process of process modeling. In [3] we discovered a relation between the order modelers construct blocks in the modeling process (structuredness) and the understandability of the resulting process model (perspicuity). Fig. 8 shows an example of a block in a model: it is a unit of alternative (OR split/join) or parallel (AND split/join) paths consisting of the split and join gateway and all elements in between. The order of creating the elements in a block can tell us more about the modeler's general modeling style. Fig. 9 shows four alternative ways of constructing a block. In Fig. 9a one can observe the modeling operations for creating the block, where the modeler created the elements from left to right. Fig. 9b shows an example where first the involved activities were created, then the gateways, and finally the edges. In the modeling process of Fig. 9c first the two activities were created, then the gateways and edges. In Fig. 9d both gateways and both nodes were created before all the edges.

The option to sort by Create order from start and the filtering of create operations enables the user to focus on this particular part of the modeling process (see Fig. 9).

## 5 Discussion and Conclusion

This paper proposes a technique for visualizing a very specific process: i.e., the process of process modeling. Supported by a modeling tool that records all modeling operations, one should be able to construct an event log that can be used as input for our technique. It graphically represents the recorded modeling operations in a so called PPMChart, which consists of time lines per modeling element and dots on these lines that represent the different operations on each modeling element in time.

The visualization technique and its implementation as a tool are based on the Dotted Chart Analysis plug-in in ProM [5]. However, we extended the functionality of this plug-in with three new features. First, we used a fixed mapping of the recorded events in the creation of a process model and the colors of the dots in the charts. This makes it easier to visually compare different charts. Second, we provided two extra sort options to match the order of the lines of a PPMChart with the order of the corresponding elements in the process model. Third, we provided filter options to hide certain model elements or modeling operations which allows a user to take different views on the data. We applied the visualization technique to the data of 103 students' processes of process modeling. To demonstrate its usefulness, we presented five concrete observed characteristic variations of the recorded modeling processes.

The main benefit of our technique is that we show raw, uninterpreted data about the process of process modeling in a way that one can quickly discover (graphical) *patterns* in the charts (that can be translated to specific properties of the modeling process). Relating these patterns to the quality of the resulting process model could help to better comprehend factors that directly influence the result of the modeling process. We would be able to utilize this knowledge in training and tools supporting process modeling, which could result in more understandable process models and a more efficient modeling process as well. At the opposite side, one must be careful to draw conclusions from the analysis of the charts, because of a lack of information about the *intentions* of the modeler. A pause in the modeling activity can indicate, for example, that the modeler was thinking about next operations or also that the modeler was distracted.

In other current research we are exploring the relation between the properties of the modeler, the way a modeler constructs process models and the quality of the resulting process model. For instance, in [3] we used the PPMChart visualization to derive three concrete conjectures about the relation of structuring, movement and speed of the modeling process to the understandability of the modeling result, which we then further tested statistically. Future research will include a search for causes and consequences of the patterns that can be discovered with PPMCharts. This will probably lead to further improvement of the visualization and its usefulness. In addition to that, we will investigate other visualization techniques that may enhance the analysis of the process of process modeling data. Currently, our visualization builds heavily on the Dotted Chart Analysis that was already available in ProM. This has given us many insights, but is not guaranteed to be the best technique. Therefore, we will investigate alternative graphical representations (such as time line trees).

**Acknowledgements**. Our research builds upon the work of the development team of CEP and the researchers involved in the experiments. Therefore, we wish to express our extensive gratitude to Stefan Zugal, Jan Mendling and Dirk Fahland. This research was funded by the Austrian Science Fund (FWF): P23699-N23.